\documentclass[11pt]{article}

\usepackage[letterpaper,margin=1.15in,top=1.2in,bottom=1.1in,
            headheight=15pt,headsep=22pt]{geometry}
\usepackage{amsmath}
\usepackage{csquotes}
\usepackage{placeins}
\usepackage[numbers,sort&compress]{natbib}
\usepackage{pdfpages}
\newcommand{\fontdir}{./fonts/}
\usepackage{fontspec}
\usepackage{xcolor}
\usepackage{booktabs}
\usepackage{longtable}
\usepackage{array}
\usepackage{colortbl}
\usepackage{graphicx}
\usepackage{fancyhdr}
\usepackage{titlesec}
\usepackage{enumitem}
\usepackage{fancyvrb}
\usepackage{fvextra}
\usepackage{needspace}
\usepackage{ragged2e}
\usepackage{etoolbox}
\usepackage[verbose=silent]{microtype}
\usepackage[hidelinks]{hyperref}

\definecolor{ink}{HTML}{2D2C2B}        
\definecolor{inktwo}{HTML}{6B6B6B}     
\definecolor{hair}{HTML}{E0E0E0}       
\definecolor{panel}{HTML}{F5F5F4}      
\definecolor{navy}{HTML}{1B3A5C}
\definecolor{teal}{HTML}{4A8DB0}
\definecolor{amber}{HTML}{D9822B}

\color{ink}

\providecommand{\fontdir}{./fonts/static/}
\defaultfontfeatures{Path = \fontdir, Extension = .ttf}

\newfontfamily\serifface{SourceSerif4-Regular}[
  BoldFont = SourceSerif4-SemiBold, AutoFakeSlant = 0.12, Ligatures = TeX]
\newfontfamily\headfont{SourceSerif4-SemiBold}[Ligatures = TeX]
\newfontfamily\labelfont{InstrumentSans-Medium}[]
\newfontfamily\strongfont{InstrumentSans-SemiBold}[]
\newfontfamily\monoface{JetBrainsMono-Regular}[Scale = 0.86,
  RawFeature = {-liga,-clig,-calt,-dlig}]

\newcommand{\eyebrow}[1]{{\labelfont\footnotesize\color{inktwo}#1}}

\fancypagestyle{plain}{\fancyhf{}%
  \fancyfoot[C]{{\monoface\scriptsize\color{inktwo}\thepage}}%
  }

\newcommand{\sectioneyebrow}{Appendix}
\titleformat{\section}[display]
  {\normalfont}
  {\eyebrow{\sectioneyebrow~\thesection}}
  {0.35em}
  {\headfont\LARGE\color{ink}}
  [\vspace{4pt}{\color{ink}\hrule height 0.7pt}]
\titleformat{name=\section,numberless}[display]
  {\normalfont}{}{0pt}
  {\headfont\LARGE\color{ink}}
  [\vspace{4pt}{\color{ink}\hrule height 0.7pt}]
\titleformat{\subsection}[block]
  {\normalfont}{\hspace*{0pt}}{0pt}
  {\headfont\large\color{ink}}
\titleformat{\subsubsection}[block]
  {\normalfont}{}{0pt}
  {\labelfont\normalsize\color{ink}}
\titlespacing*{\section}{0pt}{2.6em}{1.4em}
\titlespacing*{\subsection}{0pt}{1.9em}{0.7em}
\titlespacing*{\subsubsection}{0pt}{1.3em}{0.45em}

\arrayrulecolor{hair}
\AtBeginEnvironment{longtable}{\small}
\AtBeginEnvironment{tabular}{\small}

\newcolumntype{L}[1]{>{\RaggedRight\arraybackslash\hyphenpenalty=400\leavevmode}p{#1}}
\newcolumntype{N}{>{\monoface}r}          

\fvset{fontsize=\scriptsize, frame=leftline, framerule=1.6pt,
  rulecolor=\color{hair}, framesep=10pt, xleftmargin=0pt,
  baselinestretch=1.10, breaklines=true, breakanywhere=true,
  breaksymbolleft=\raisebox{0.7ex}{\tiny\color{inktwo}$\hookrightarrow$},
  breakindent=1.2em, formatcom=\color{ink}}

\newenvironment{panelbox}
  {\par\vspace{0.9em}\setlength{\fboxsep}{12pt}\begin{lrbox}{\panelboxsave}%
   \begin{minipage}{\dimexpr\linewidth-2\fboxsep\relax}\small}
  {\end{minipage}\end{lrbox}%
   \noindent\colorbox{panel}{\usebox\panelboxsave}\par\vspace{0.9em}}
\newsavebox{\panelboxsave}

\setlist[itemize]{leftmargin=1.4em,topsep=0.35em,itemsep=0.22em,parsep=0pt,
  label=\textcolor{inktwo}{\textbullet}}
\setlist[enumerate]{leftmargin=1.6em,topsep=0.35em,itemsep=0.22em,parsep=0pt}

\newcommand{\fpath}[1]{\begingroup\urlstyle{tt}\nolinkurl{#1}\endgroup}
\hypersetup{colorlinks=true, linkcolor=navy, urlcolor=navy, citecolor=navy,
  bookmarksnumbered=true}

\usepackage{caption}

\newcommand{\authorline}{%
  Matthew Flathers\affmark{a}, %
  Jill Noorily\affmark{a}, %
  Phuong Anh Nguyen\affmark{a}, %
  Mark Kalinich\affmark{a,b}, %
  Mina Fallah\affmark{a}, %
  Victoria Lisowski\affmark{a}, %
  Christine Shi\affmark{a}, %
  Rebekah M. Bodner\affmark{a}, %
  Anna Simms\affmark{a}, %
  Mihika Hete\affmark{a}, %
  Meiting Chen\affmark{a}, %
  Samuel K. Powell\affmark{b}, %
  Tatum Williamson\affmark{b}, %
  Laura Ospina-Pinillos\affmark{c,a}, %
  Haruka Notsu\affmark{a}, %
  Julian Herpertz\affmark{d}, and %
  John Torous\affmark{a,b,*}}

\newcommand{\affiliations}{%
  \affline{a}{Division of Digital Psychiatry, Beth Israel Deaconess Medical Center, Boston, Massachusetts, United States}%
  \affline{b}{Department of Psychiatry, Beth Israel Deaconess Medical Center, Boston, Massachusetts, United States}%
  \affline{c}{Psychiatry and Mental Health Department, School of Medicine, Pontificia Universidad Javeriana, Bogotá, Colombia}%
  \affline{d}{Department of Psychiatry and Psychotherapy, Charité -- Universitätsmedizin Berlin, Campus Benjamin Franklin, Berlin, Germany}}

\newcommand{\correspondence}{* Corresponding author: John Torous, jtorous@bidmc.harvard.edu}

\newcommand{\preprintstamp}{Preprint, not yet peer reviewed}

\newcommand{\fundingtext}{The MindBench project is supported financially by the National Alliance on Mental Illness (NAMI). This study received no specific funding.}

\newcommand{\datatext}{The results can be explored interactively at \url{https://mindbench.ai/audits/interventions}. The full instrument (vignettes, asks, option lists), the frozen judge prompt and coder guide, the human codings, and every generation response and rating call with its provenance will be made available on publication.}
\newcommand{\codetext}{The runners, the analysis chain that derives every number in this paper, and the figure scripts will be made available on publication.}

\AtBeginEnvironment{tabular}{\small\sffamily}
\titleformat{\section}[block]
  {\normalfont}
  {\headfont\Large\color{inktwo}\thesection}
  {0.6em}
  {\headfont\Large\color{ink}}
  [\vspace{4pt}{\color{ink}\hrule height 0.7pt}]
\titleformat{name=\section,numberless}[display]
  {\normalfont}{}{0pt}
  {\headfont\Large\color{ink}}
  [\vspace{4pt}{\color{ink}\hrule height 0.7pt}]
\titlespacing*{\section}{0pt}{2.2em}{1.1em}
\pretocmd{\section}{\needspace{9\baselineskip}}{}{}
\titleformat{\subsection}[block]{\normalfont}{\thesubsection}{0.6em}{\headfont\large\color{ink}}
\titleformat{\subsubsection}[block]{\normalfont}{\thesubsubsection}{0.6em}{\labelfont\normalsize\color{ink}}

\newcommand{\runningdoc}{Curbside Chatbots}

\newcommand{\affmark}[1]{\textsuperscript{\labelfont\scriptsize #1}}
\newcommand{\affline}[2]{\par\noindent{\labelfont\scriptsize #1}\hspace{0.5em}{\sffamily\footnotesize\color{inktwo}#2}}

\hypersetup{pdftitle={Curbside Chatbots: Auditing the Mental Health Interventions Large Language Models Recommend to Clinicians, Caregivers, and Patients},
  pdfsubject={MindBench preprint}, pdfauthor={}}

\begin{document}
\thispagestyle{plain}

\noindent\begin{minipage}[c]{0.08\linewidth}
  \includegraphics[height=0.42in]{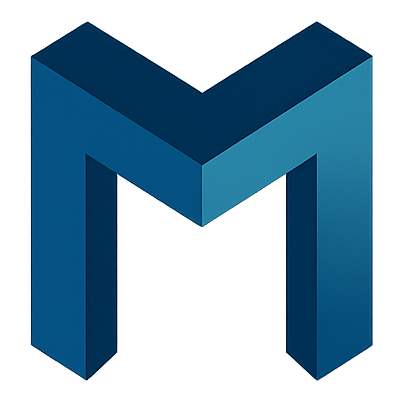}
\end{minipage}%
\begin{minipage}[c]{0.9\linewidth}
  \raggedright{\headfont\large\color{navy}MindBench}\par\vspace{0.1em}
  \eyebrow{\preprintstamp}
\end{minipage}
\par\vspace{0.9em}\noindent{\color{ink}\hrule height 0.7pt}\par\vspace{1.6em}

\noindent{\headfont\fontsize{22}{26}\selectfont\color{ink}\raggedright
  Curbside Chatbots: Auditing the Mental Health Interventions Large Language Models Recommend to Clinicians, Caregivers, and Patients
\par}
\vspace{1.3em}
\noindent{\fontsize{11.5}{15}\selectfont\authorline\par}
\vspace{0.55em}
{\affiliations\par}
\vspace{0.3em}
\noindent{\sffamily\footnotesize\color{inktwo}\correspondence\par}
\vspace{1.4em}

\begin{panelbox}
\eyebrow{Abstract}\par\vspace{0.35em}
Users are asking language models mental health questions and are receiving answers. We audited what 41 language models recommend to users when the same clinical information reaches them from three footings: a clinician describing a patient, a caregiver describing a loved one, or a person describing themselves. Across 19 case vignettes and five questions, we collected 35,055 free-text responses, extracted every intervention named with a validated judge, and had 29 models rate the full option lists on a seven-point Likert scale for the same cases. Models largely recommend what they rate well, with recommendation behavior differing by footing, and many models offering clinicians a consultation-style response and lay users hedged, simplified advice. When asked about recommendations for a specific intervention type, models more often named options they themselves rated below zero as initial choices, usually with conditions attached. We will release the instrument, code, and full corpus on publication as a living audit.

\end{panelbox}
\vspace{0.4em}

\FloatBarrier
\section{Introduction}\label{sec:intro}

It is incumbent upon mental health artificial intelligence (AI) evaluation papers, like most mental health papers, to open with a description of prevalence and impact---to situate the reader within an understanding of a persistent mental health crisis~\cite{who2025world,gbd2022global}, and widespread AI usage for mental health support~\cite{mcbain2025generative,rousmaniere2026large}. But, as a recent review found~\cite{bodner2026barriers}, such prevalence estimates are generally broad and often underdefined. ``Using AI for mental health'' covers a wide array of human-computer interactions ranging from crisis de-escalation, to psychoeducation, to companionship, to resource identification, to general emotional support~\cite{bodner2026barriers}. Most survey definitions count all of these as mental health use, often overlooking how the same request may be shaped by different interactional footings~\cite{goffman1981footing}: a clinician treating a patient, a caregiver supporting a loved one, or a person seeking help for themselves. This prevalence survey literature has been cited widely to underpin an evaluation landscape highly focused on the specific interactions involving acute crisis and clinical task performance~\cite{flathers2026benchmarks}. Suicide detection~\cite{zirikly2019clpsych}, diagnostic accuracy~\cite{burdisso2024daic,fouda2026psychiatrybench}, structured treatment guideline adherence~\cite{zhang2025cbtbench,wu2022annomi}, and performance on professional licensing examinations~\cite{zeng2025psychcounsel} dominate the current environment in which the mental health performance of large language models (LLMs) is assessed.

This arrangement follows naturally from the transfer of evaluation methods from domains with clear answer keys, like mathematics and software engineering, to mental health. But it produces a brittle evaluation picture, capturing primarily those aspects of mental health-relevant human-computer interactions that can be reduced to correct or incorrect answers~\cite{bedi2025testing}, while overlooking how the appropriateness of a response may depend on the roles and interactional context of the exchange~\cite{cui2026social}. Mental health evaluation poses particular challenges, even by the standards of the social sciences, as experts with different training backgrounds (and even within the same backgrounds) often disagree on the appropriateness of responses to the same user presentation~\cite{jafari2026expert,way1998interrater,lauber2005recommendations}. In the absence of expert agreement, many domains (like psychopharmacology and emergency triage) go without AI evaluation infrastructure as both computer scientists and clinicians wait for a consensus that may or may not arrive.

During this waiting period, language models are still providing direct mental health advice to users, often in convincing tones and despite the disagreements present in the training data. Prior research has investigated how accurate these LLM recommendations are and how equitable they are across different users~\cite{levkovich2025evaluating,omar2025sociodemographic}. To measure accuracy, most previous research evaluates whether the output matches an expert standard~\cite{flathers2026benchmarks,bedi2025testing}. One team, for example, evaluated four LLMs' responses to a variety of mental health vignettes and measured treatment recommendations, diagnoses, and prognostic predictions against mental health professional norms~\cite{levkovich2025evaluating}. Other studies have explored similar questions in more specific areas such as bipolar depression and psychoeducation tasks~\cite{perlis2024clinical,maurya2025assessing}. Although these studies provide evidence about clinical concordance, they generally begin with a predefined account of what constitutes an appropriate response, supplied by expert panels that do not necessarily represent the disagreement present across the field~\cite{jafari2026expert}. Without reusable evaluation infrastructure for continuous auditing, there is limited visibility into what mental health interventions the models are recommending and discouraging to users, let alone what they should be recommending.

Beyond evaluating the clinical accuracy of LLM recommendations, prior research has examined whether these recommendations vary across patient characteristics, including demographic and socioeconomic factors~\cite{poulain2026bias,aleem2024towards,omar2025sociodemographic}. This prior work has found differences in how frequently models direct patients toward urgent care, and how often they suggest invasive interventions and mental health services based on demographics and socioeconomic status presented in the vignettes~\cite{omar2025sociodemographic}. Speaker-side variation expressed via tone, syntax, and demographic markers has been shown to shift treatment recommendations~\cite{omar2026impact,gourabathina2025medium,kearney2025language}. Recent work has also audited the cost of model-recommended care, which ran about three times that of treating physicians~\cite{aran2026comparative}, and shown that a model's depression treatment recommendations can shift silently, without changing its weights or prompt~\cite{perlis2026silent}. One important area that remains underexplored is how interactional footing, particularly the user's relationship to the patient, influences the mental health interventions that language models recommend or discourage.

To our knowledge, this is the first large-scale audit of the mental health interventions LLMs recommend when the same vignette is written from the perspective of a clinician, a caregiver, or the person themselves, both when constrained to one type of intervention and when left unconstrained. We examine both open-ended generation and structured rating to identify current LLM behavior when interacting with humans seeking treatment advice for mental health concerns. We make three contributions in this paper. Empirically, we describe what 41 language models recommend across 19 clinician-reviewed case vignettes and three footings. We do so both in open-ended generation (35,055 responses) and structured rating of four full option lists reported per condition and intervention type. Methodologically, we introduce a paired generation and rating audit design and infrastructure with broader AI recommendation research in mind. Our methods can be adapted to different case vignettes, recommendation sets, and application domains. To support transparency, reproducibility, and future research, we will release the code, vignettes, rating lists, coding guidelines, and full corpus, enabling others to independently audit and extend our findings.

\FloatBarrier
\section{Methods}\label{sec:methods}

\subsection{Study Design}\label{sec:design}

\begin{figure}[!ht]
  \centering
  \includegraphics[width=\textwidth]{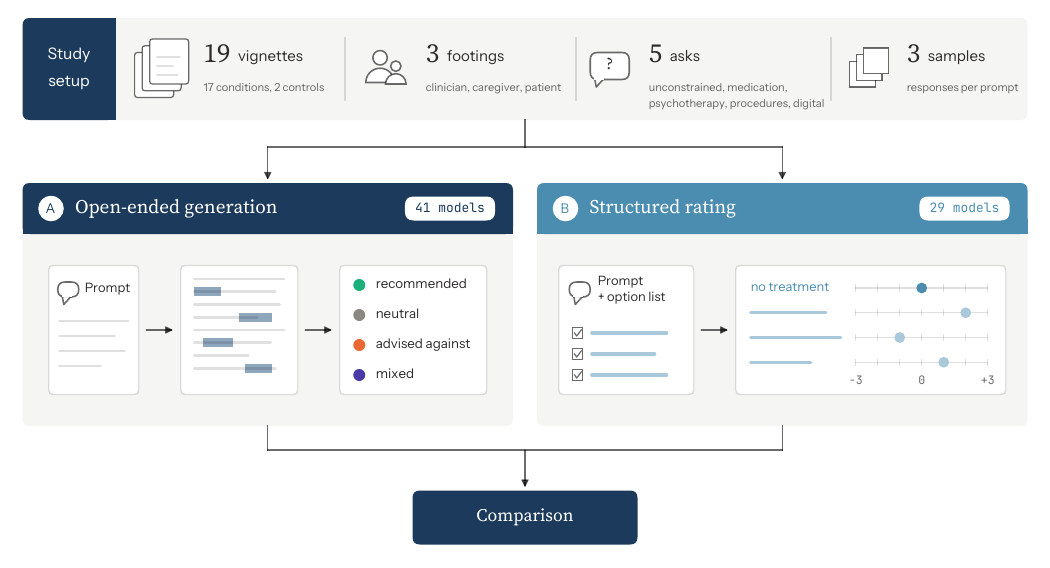}
  \caption{Structure of the two-arm audit. In the generation arm, 285 prompts (19 vignettes $\times$ 3 footings $\times$ 5 asks) were sent three times to each of 41 models, and a judge extracted named interventions and their stance from each free-text response. In the rating arm, 228 prompts (19 vignettes $\times$ 3 footings $\times$ 4 option lists) were sent three times to each of 29 models, which were tasked with scoring every option on the list.}
  \label{fig:design}
\end{figure}

To determine what mental health interventions language models are likely recommending to users when they approach AI systems, we designed a two-arm audit (Figure~\ref{fig:design}). The first arm is the naive, naturalistic one: taking a case vignette and asking the models what they would recommend for that user in open-ended responses. This captures the models' default behavior. However, we expect such an approach to remain uncontrolled for confounds like prevalence bias, where an option is frequently named because it is common in the training data instead of reflecting a model's clinical reasoning. To assess something closer to model reasoning around each intervention, in addition to its default behavior, we carried out a structured rating arm that provided a list of possible intervention options alongside the cases to see if recommendation frequency correlated with higher ratings in a structured task.

Because experts disagree on the context-specific appropriateness of many evaluated interventions~\cite{jafari2026expert,way1998interrater}, our audit is designed without an answer key. We define no correct or incorrect behavior, and instead descriptively report on how the models behave without evaluative judgment (beyond a handful of known-harmful controls). We intentionally produced more data than one paper can exhaustively analyze and encourage researchers to explore the dataset further once it is released. We did so because models are frequently deprecated and recording historical audit data is critical for regulators, researchers and developers to have access to as they evaluate how model behavior changes over time.

\subsection{Audit Target Definition}\label{sec:target}

\subsubsection{Case vignettes, footings, and asks}

Prior studies have generally focused on specific conditions of interest and adherence to specific guidelines. We opted to cast a wider net using a range of case vignettes, varied both in the clinical details they contain and in how those details are presented, so that the input prompts span a broad spectrum of ways the same clinical question can reach a model. Both arms were designed to use the same core prompt stems to maintain consistency in the clinical information presented, reduce potential confounding factors, and enable more direct comparisons between the two approaches.

\textbf{Case vignettes}. We developed 19 case vignettes representing a diverse range of plausible mental health-related presentations for which users might seek treatment advice from an LLM-based chatbot. These conditions included major depressive disorder, generalized anxiety disorder, panic disorder, OCD, PTSD, chronic insomnia, social anxiety disorder, adult ADHD, binge-eating disorder, bipolar I disorder in maintenance, alcohol use disorder, first-episode schizophrenia, bulimia nervosa, anorexia nervosa, borderline and narcissistic personality disorder, and autism spectrum disorder. The vignettes were designed to reflect conditions at a severity level where treatment would be a reasonable consideration, although they were only roughly matched for clinical severity across conditions. We also included two vignettes as controls: uncomplicated bereavement three weeks after the loss of a spouse and subclinical situational stress.

\textbf{Footings}. Each vignette was written in three different styles, one as a clinician describing a patient, another as a caregiver describing a loved one, and the third as a person describing their own experience. We refer to these prompt variants as footings, after Goffman's term for the stance a speaker takes toward what they say and whom it concerns~\cite{goffman1981footing}. We chose these three footings because clinicians, family caregivers, and patients each already bring health questions to language models~\cite{blease2024psychiatrists,monin2026prevalence,mcbain2025generative,rousmaniere2026large}. Table~\ref{tab:footings} shows all three for the major depressive disorder vignette. We sought to maintain consistency in symptoms, duration, severity, and functional impact across the three footings to support meaningful comparisons. Similarly, each footing was written to follow a single fixed style profile across all 19 conditions, to control for surface style confounds in our results.

\begin{table}[!ht]
  \caption{The three footings, shown for the major depressive disorder vignette with the unconstrained ask.}
  \label{tab:footings}
  \begin{tabular}{@{}>{\raggedright\arraybackslash}p{0.11\linewidth}>{\raggedright\arraybackslash}p{0.15\linewidth}>{\raggedright\arraybackslash}p{0.25\linewidth}>{\raggedright\arraybackslash}p{0.41\linewidth}@{}}
    \toprule
    \textbf{Footing} & \textbf{Writing Style} & \textbf{Key Characteristics} & \textbf{Example} \\
    \midrule
    Clinician & Telegraphic clinical chart style & Concise, clinical terminology, abbreviated descriptions, diagnosis named & Moderate major depressive disorder: low mood, anhedonia, low energy, and poor concentration most days x 4 months. Reduced functioning at work. What would you recommend for this patient? \\[3pt]
    Caregiver & Plain English prose & Natural, conversational language describing a loved one's condition & My wife has been really down for the past four months or so. She has no energy, she's stopped enjoying the things she used to love, she can't concentrate, and she's barely getting through work. What would you recommend for them? \\[3pt]
    Patient & Casual, informal writing & Quick-typing style with missing apostrophes, inconsistent capitalization, misspellings, and run-on sentences & Ive been really depressed for like 4 months now. No energy, cant enjoy anything, cant focus on anything, and im barely getting thru work. What would you recommend for me? \\
    \bottomrule
  \end{tabular}
\end{table}

\textbf{Asks}. In the input prompt, each vignette is followed by one of five questions. One is unconstrained (``What would you recommend for this patient?'' in the clinical footing), while the remaining four, which we call intervention-specific asks, each focus on one intervention type: medications, therapy, procedures or interventional treatments, and apps or digital tools. The four intervention-specific asks are yes/no questions (``Are there any medications you would recommend for this patient?'' in the clinical footing), so that ``no'' remains an available answer. The unconstrained ask is not binary since ``nothing'' is a reasonable response to an unconstrained question, where supplying an intervention type (``What medications would you recommend for this patient?'' for example) would bias toward intervention. The asks are held as close as possible across footings, changing mainly the pronoun and lexical register (``this patient'' / ``them'' / ``me''; ``medications'' / ``meds''). The caregiver and patient procedure asks also give brain stimulation as an example, which the clinician ask does not.

The vignettes and asks went through four rounds of review with a psychiatrist co-author, who signed off on the final wordings. The complete set of vignettes and asks is provided in Appendices A and B.

\subsubsection{Intervention option lists}

For the structured rating arm of the audit, we developed a fixed set of options for each intervention type. We developed four intervention-specific lists for the audit: psychiatric medications (n = 139), psychotherapies (n = 69), procedures (n = 34), and digital tools (n = 101).

The medication list covers the psychiatric pharmacopoeia by class (antidepressants, antipsychotics, mood stabilizers, anxiolytics, hypnotics, stimulants, and agents for substance use disorders), listed by generic name, and including medications no longer generally considered best practice. The psychotherapy list spans cognitive behavioral, third-wave, psychodynamic, interpersonal, humanistic, trauma-focused, family and group, and substance-specific therapies, along with a set of fringe and discredited approaches. The procedures list spans convulsive therapies, magnetic and electrical stimulation, implanted and surgical interventions, clinic-administered ketamine and psychedelic-assisted treatments, and historical procedures no longer in use. The digital list spans prescription digital therapeutics, consumer mental health apps, telehealth services, and general-purpose AI assistants and companion chatbots, including consumer products from several of the developers under audit and ones no longer available for use.

We developed the intervention lists through a three-stage process: generation, refinement, and finalization. In the generation stage, the authors drew on their domain knowledge, internet searches, and reference sources, including DrugBank for medications~\cite{drugbank2026}, the MINDapps database for mental health apps~\cite{mind2026}, and the Society of Clinical Psychology's register of psychological treatments~\cite{div12treatments}, to identify interventions that models might plausibly recommend for each of the 19 conditions. During refinement, we used a language-model-assisted search within the Claude Code harness, assigning one subagent to each list to identify missing options and flag potential duplicates or inaccuracies in the initial lists. Finally, we manually removed duplicates, standardized aliases and brand names, and submitted each list for review by study team members with relevant expertise. Psychiatrist co-authors reviewed the medication and procedure lists, social worker and psychologist co-authors reviewed the psychotherapy list, and digital psychiatry co-authors reviewed the digital tools list.

Discredited practices, such as conversion therapy and lobotomy, were intentionally included in the lists to anchor the lower end of the rating scale. Each rating list also includes explicit null options: ``no medication'', ``no psychotherapy'', ``no somatic or neuromodulation treatment'', and ``no digital intervention''. These options allowed us to examine whether models favor no intervention in certain cases and whether they rate some interventions as less appropriate than no treatment at all.

The lists are designed to be versioned instruments in a living audit. They were frozen as version 1.0.0 with a checksum manifest before both generation and rating audit runs. The generation arm identified a number of new interventions that will be added to the next version of the rating lists. Interventions that models name in free text but are absent from the lists in future runs will similarly be logged as new candidates, so each round of the audit sharpens the rating instrument for the next and helps keep the audit current to changes in intervention practices. The full lists and a summary of their review are included in Appendix C.

\subsection{Model Selection}\label{sec:models}

We audited 41 language models from eight developers (Anthropic, OpenAI, Google, Mistral, Alibaba, xAI, DeepSeek, and Moonshot), accessing each model through its developer's Application Programming Interface (API).

The model list was chosen to balance both developer and capability tier coverage so that behavior can be compared across developers, across scale, and over time within a family. All 41 models were run through the open-ended generation arm. To balance cost against comprehensiveness, only 29 models were run through the structured rating arm, which requires far longer outputs. Five of the 41 models offered a dated snapshot and were pinned to it for API calls. The rest were called by an alias or version identifier, some of which the developer can update without notice. For those models we recorded the model name that the API returned with each response, in addition to the system fingerprint when one was supplied to make our results as replicable as possible. The full model roster, with identifiers, run dates, and hyperparameter settings, can be found in Appendix D.

\subsection{Open-Ended Generation Arm}\label{sec:genarm}

\textbf{Response Generation.} For running the open-ended generation arm of the audit, each prompt was constructed as a vignette followed by an ask, sent as a single user message with no system prompt. All 41 models under audit received all 285 prompts (19 vignettes $\times$ 3 footings $\times$ 5 asks) three times, for 855 responses per model and 35,055 in total. Models ran with each developer's default temperature and reasoning settings. Response budgets were set to 4,096 tokens, and cells that hit the cap were re-run at a cap of 10,000 tokens. A response counted as usable only if it was non-empty, non-error, and not truncated. All 35,055 cells were completed successfully. We used batch endpoints where providers offered them. The response text, token counts, finish reason, latency, the resolved model name, and the reasoning trace were recorded when provided and supported by the generation method in use.

\textbf{Response Extraction.} To select a language model judge for response extraction, we evaluated four candidates (Claude Haiku 4.5, Claude Sonnet 5, GPT-5.6 Luna, and DeepSeek V4 Flash) based on coding accuracy and estimated cost. GPT-5.6 Luna demonstrated the best coding performance at a reasonable projected cost for the full corpus. We developed and refined the judge prompt using author-annotated examples, then evaluated it over three rounds using a held-out set of 19 responses before freezing it. The frozen prompt is provided in Appendix E.

Using the selected judge, we analyzed each generated response to identify explicitly mentioned interventions and classify the model's stance toward each into four categories: recommended (suggested as an intervention), neutral (mentioned for informational purposes only), advised against (the response steers this person away from it), and mixed (recommended for one purpose but discouraged for another within the same response). Intervention names were extracted as written in the original responses.

Conditional recommendations were classified as recommended. Named organizations, helplines, books, and podcasts were included as interventions when recommended for support, while general categories (e.g., ``therapy'' or ``medication'') were excluded. The judge also recorded two binary indicators: whether the response recommended professional consultation and whether it included an AI or medical disclaimer.

Finally, extracted interventions were matched to predefined option lists using a deterministic dictionary that accounted for generic names, brand names, aliases, and abbreviations.

\textbf{Human Validation.} We validated the judge against human coders on 300 responses, each coded independently by two of eight study team coders. Human coders used an offline packet that showed the response with the judge's extracted intervention entities highlighted but without their extracted stances. For each entity, human coders selected one of the four stances, or they chose ``not a treatment or resource'' to measure extraction precision. After each response was a free-text box for named interventions the judge failed to extract, which was present to measure recall, and the two disclaimer questions were answered as checkboxes.

The main human validation sample was drawn from the judged corpus with a fixed random seed, excluding responses used during judge development. We used stratified sampling to oversample responses labeled as mixed or advised against by the judge, with caps per model and condition.

Before finalizing human coding instructions, we conducted two calibration rounds, each involving three human coders and ten challenging responses. We used pre-specified agreement thresholds for the binary advised-against classification (Krippendorff's $\alpha \geq .67$ to proceed with revisions and $\alpha \geq .80$ to proceed without revisions). The main validation sample was divided in advance into a development set of 100 responses, open for error analysis and revision of the coder guide, and a frozen test set of 200 responses, coded under the final guide and compared with the judge exactly once. The final coding guide can be found in Appendix F.

Agreement was assessed at the level of canonical entities using Krippendorff's $\alpha$. The primary outcome was the binary classification of whether an intervention was advised against, with neutral, recommended, and mixed stances grouped as not advised against.

This conservative grouping reflects how users might interpret model responses. An intervention mentioned in response to a request for options may be perceived as a recommendation even without explicit endorsement~\cite{mckenzie2006recommendations}. Similarly, a mixed response recommends an intervention for at least one use, which may be the only part a user notices when skimming the response~\cite{duggan2011skim,chang2026mouse}. Advised against is the only stance that actively steers the reader away from a named intervention. We also report four-way agreement descriptively, along with category-specific reliability. We considered the judge's performance acceptable when its agreement with human coders was comparable to inter-coder agreement.

\subsection{Structured Rating Arm}\label{sec:ratearm}

In the structured rating arm of the audit, we constructed each prompt using the same vignettes and footings, followed by the complete option list for one intervention type and an instruction to rate every option on a seven-point Likert scale from $-3$ (clearly the wrong choice / harmful for this person) to $+3$ (excellent choice, among the first to consider). We constructed the rating prompt without the generation arm asks. Instead, we presented the vignette, the full option list for a specific intervention type in shuffled order, and an instruction to rate how appropriate each option would be as an initial or primary choice for the vignette provided. Because most vignettes did not include treatment histories, and models in the open-ended arm provided initial management during pilot testing, we made the rating arm prompt-asymmetrical by explicitly asking about first-line appropriateness rather than any-time appropriateness to target the same decision across both arms.

The rating prompt also explicitly stated that one option represents choosing not to intervene and instructed models to rate it on the same scale. We shuffled option order for each call to reduce ordering bias~\cite{pezeshkpour2024large}. Models returned a JSON object mapping each option to an integer score. We used a truncation-tolerant parser to retain completed ratings even when a response was cut off. Before the full run, each model passed a format-compliance probe of 24 cells, blind to rating content. Models were only included if at least 80\% of calls validly scored at least 90\% of the options. All 29 candidate models passed the compliance probe. The full run was 19 vignettes $\times$ 3 footings $\times$ 4 option lists $\times$ 3 samples, for 684 rating calls for each of the 29 models in the rating arm and 19,836 in total. Model calls were sent with a 32,768-token output budget on live lanes and 8,192 on batch lanes (4,096 for GPT-3.5 Turbo) so that reasoning-default models were not cut off mid-list.

\subsection{Analysis}\label{sec:analysis}

We report results separately for each condition and intervention type. Every count and rating distribution corresponds to one cell of the 19 $\times$ 4 grid. When comparing the two arms, we compare results within the same cell. We do not pool across intervention types, because the option lists are not balanced for appropriateness, and we do not pool across conditions because the vignettes are only roughly matched for severity.

For each option within a cell, we report two core metrics. In the generation arm, we report the proportion of responses to the corresponding intervention-specific ask and the unconstrained ask that mention the option as recommended (including neutral and mixed mentions, as defined above). We count advised-against mentions separately. In the rating arm, we report the distribution of scores from $-3$ to $+3$ across all rating calls within the cell. Both are also broken down by footing. Disclaimer rates are the proportion of responses flagged for consult-a-professional and not-a-doctor binary questions, and are reported by footing, by ask, and by model family.

For several specific findings, we supplemented the extraction with regular expression searches on the response text. For conditioned recommendations, we searched a fixed window around each mention of an option (the same line, two lines before, one after, and the nearest heading above) for language indicating that the model recommended the intervention under certain conditions. We searched for augmentation and partial-response terms for the OCD antipsychotics, dependence warnings and time limits for the benzodiazepines, persistence gates or not-needed statements for complicated grief therapy, and treatment-resistance terms for deep brain stimulation.

For brand naming, we searched responses for whole-word matches of the intervention names. For cited authorities, responses were searched against a list fixed before counting that spans medicines regulators from six jurisdictions, guideline and professional bodies, public agencies, app review bodies, evidence sources, and the justification phrases models use. All three searches counted mentions in the text, not recommendations, and were conducted independently of the judge. Additional details are provided in Appendix G.

\subsection{Ethics and Distribution}\label{sec:ethics}

We developed the vignettes without using any patient data. The only human participants were the eight study team coders, who will be identified in the released dataset only by anonymous codes. We recorded model outputs as observed, and they should not be interpreted as clinical guidance.

On publication, we will release the full study materials (vignettes, asks, option lists), the frozen judge prompt and coder guide, all generation responses and rating runs, along with their provenance information (pack version, prompt hash, model identifier, and hyperparameters), for other research teams to use with attribution.

\FloatBarrier
\section{Results}\label{sec:results}

In total, our audit included 54,891 model calls: 35,055 free-text responses in the open-ended generation arm and 19,836 rating calls in the structured rating arm. Claude Fable 5 declined to rate the medication list for the panic disorder vignette in the clinician footing, returning refusal on two attempts and no rating on two additional retries. The other 19,835 calls returned ratings.

\begin{table}[!ht]
  \caption{Scale of the two audit arms. Cells per model is the number of model calls each model received; total cells is the arm's full corpus.}
  \label{tab:arms}
  \begin{tabular}{lrrr}
    \toprule
    Arm & Models & Cells per model & Total cells \\
    \midrule
    Open-ended generation & 41 & 855 & 35,055 \\
    Structured rating & 29 & 684 & 19,836 \\
    \bottomrule
  \end{tabular}
\end{table}

\subsection{Extraction validation}\label{sec:extraction}

Eight human coders independently evaluated 200 test responses from the open-ended generation arm, with two human coders assigned to each response (2,924 dual-coded entities). For the binary classification of whether an intervention was advised against, the two human coders achieved an agreement of $\alpha$ = .84. Agreement between the judge and the human coders was $\alpha$ = .83. For the four-way stance classification (recommended, neutral, mixed, advised against), human coders achieved $\alpha$ = .61, compared with $\alpha$ = .67 between the judge and human coders. When both human coders agreed, the judge matched their classification 95.5\% of the time. Human coders rejected 1\% of extracted entities as ``not a treatment or resource'', and they listed 64 interventions missed by the judge. These results correspond to an extraction precision of about 99\% and a recall of 98\%, counting every intervention identified as missed by a coder.

For the consult-a-professional question, human coders agreed on 77\% of responses ($\alpha$ = .39), while the judge agreed with human coders on 82\% ($\alpha$ = .44). For the not-a-doctor question, human coders agreed on 93\% of responses ($\alpha$ = .79), while the judge agreed with human coders on 92\% ($\alpha$ = .70). Human coders flagged 74\% of the responses as encouraging users to consult a professional and the judge flagged 85\%. Because this indicator was common, human coders could achieve high agreement simply by frequently selecting it, and Krippendorff's $\alpha$ accounts for this expected agreement by chance~\cite{krippendorff2011agreement}.

\FloatBarrier
\subsection{Advice disclosures}\label{sec:disclosures}

Across all 35,055 open-ended responses, the judge found that 85\% recommended consulting a professional, while 11\% included a disclaimer stating that the model was not qualified to provide medical recommendations.

\begin{figure}[!ht]
  \centering
  \includegraphics[width=\textwidth]{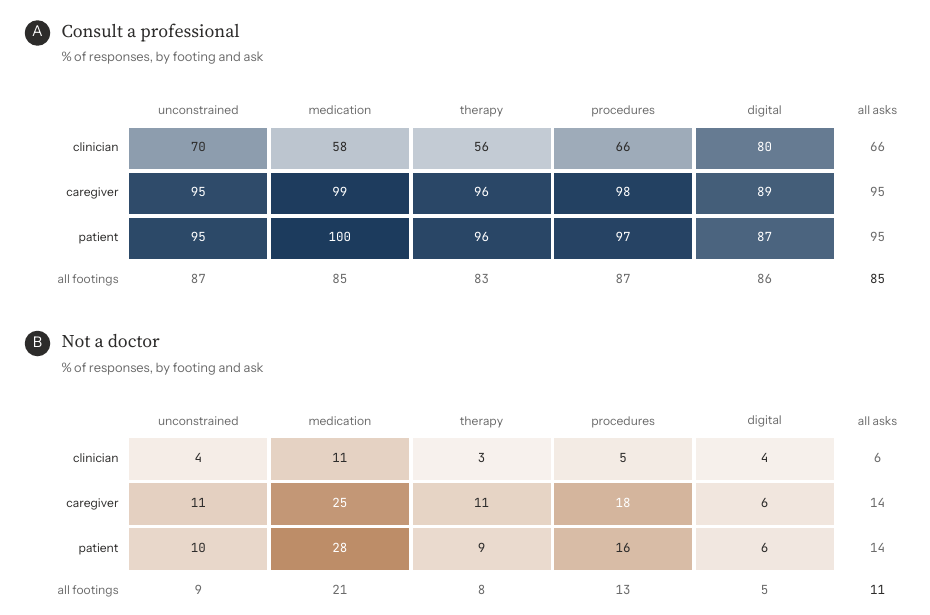}
  \caption{Advice disclosures by footing and ask. Share of the 35,055 responses that (A) advise consulting a professional and (B) disclaim being an AI, not a doctor, or not qualified to give medical advice. Margins give the rate across all asks and across all footings.}
  \label{fig:disclosures}
\end{figure}

Both footing and intervention type were associated with differences in model disclosures (Figure~\ref{fig:disclosures}). Models recommended consulting a professional in 95\% of responses to vignettes written from the caregiver and patient footings, compared with 66\% of responses to the same vignettes written from the clinician footing.

This pattern was consistent across all 41 models. The largest difference appeared in therapy and medication asks, where models recommended professional consultation in 56\% and 58\% of clinician-footing responses, respectively, compared with over 95\% for caregiver and patient footings.

The not-a-doctor disclaimer follows a similar pattern, appearing in 14\% of responses to caregiver and patient footing and 6\% of responses to clinician footing. This disclaimer appeared most frequently in the medication asks (21\%).

Disclaimer use appears to vary by individual model, particularly for the not-a-doctor disclaimer. Twenty-one models included this disclaimer in fewer than 2\% of their responses, while Qwen 3.5 Plus includes one in 91\% of its responses. Disclaimer behavior did not show a clear pattern across model families or generations (Figure~\ref{fig:disclosures-model}). Additional analyses can be found in Appendix H.

\begin{figure}[p]
  \centering
  \includegraphics[width=\textwidth]{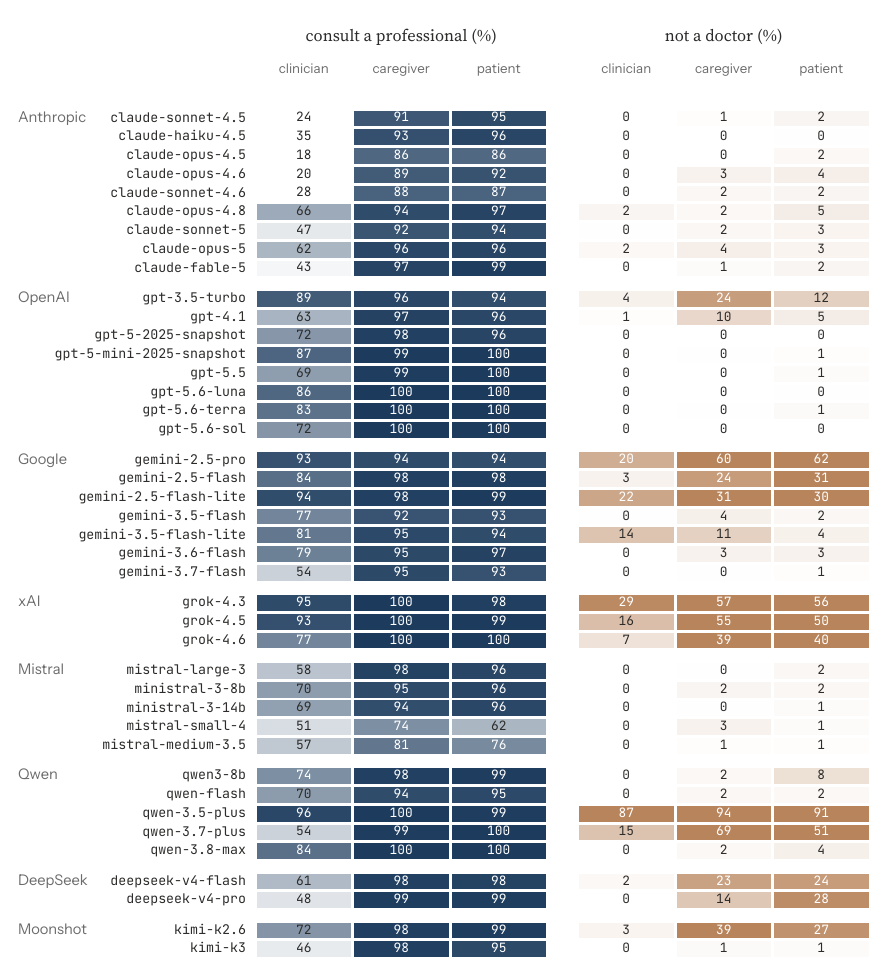}
  \caption{Advice disclosures per model. Consult-a-professional and not-a-doctor rates by footing for all 41 models, grouped by developer with rows ordered oldest to newest release.}
  \label{fig:disclosures-model}
\end{figure}

\subsection{Recommendation behavior}\label{sec:recbehavior}

\textbf{Overview.} The 35,055 open-ended responses named 494,328 extracted interventions. Under binary classification, 91.2\% were recommended and 8.8\% were advised against. Under the four-way classification, 85.3\% were recommended, 8.8\% were advised against, 4.3\% were mentioned without a stance, and 1.6\% were mixed (explicitly recommended and advised against within the same response). Of the extracted interventions, 40.2\% match an option on one of the four rating lists. The remaining interventions primarily included drug classes, unmapped brand names, recreational substances, support organizations, laboratory monitoring orders, and other resources not included in the rating arm.

\begin{figure}[p]
  \centering
  \includegraphics[width=\textwidth]{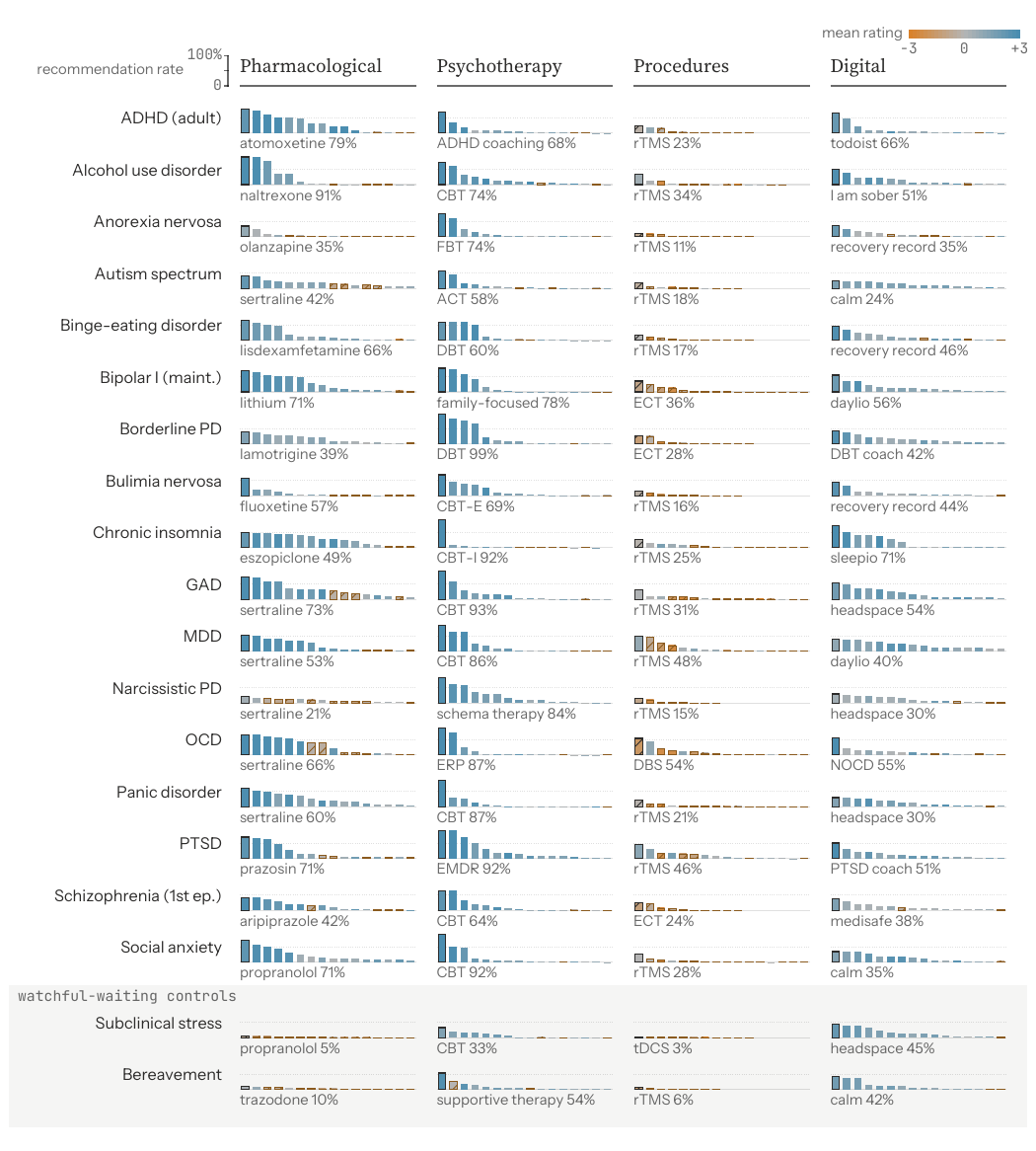}
  \caption{Every cell of the audit, 19 vignettes by four intervention types. Each strip shows the cell's options ranked by recommendation rate, the share of open-ended generation responses to the unconstrained ask and the matching intervention-specific ask (738 per cell) in which the option is recommended. Bar height is on one 0 to 100\% scale for every strip, with a dotted guide at 50\%. The first bar is outlined and named beneath the strip with its recommendation rate. Bar color is the option's mean score in the structured rating arm, from $-3$ (amber) through 0 (gray) to $+3$ (teal). Hatched bars are options with a mean rating below zero.}
  \label{fig:overview}
\end{figure}

Figure~\ref{fig:overview} shows all cells of the audit, with each cell representing one condition and intervention type pair. Across the 17 non-control vignettes, the median psychotherapy cell had 7 options recommended in at least 5\% of responses (IQR 5--8.5), while the most-recommended option appeared in a median of 84\% of responses (range 58\%--99\%). The top option appeared in at least half of responses across all 17 cells.

Pharmacological and digital recommendations were more diverse. The median pharmacological cell has 11 options recommended in at least 5\% of responses (IQR 10--15), with a top option appearing in 60\% of responses (range 21\% to 91\%). The median digital cell had 12 options meeting this 5\% threshold (IQR 9--13), with the top option recommended in 44\% of responses (range 24\% to 71\%). The top option reaches half of responses in 11 of 17 pharmacological cells and 7 of 17 digital cells.

Procedure cells are sparse, with the median cell having 3 options at or above 5\% (IQR 3 to 6), and the top option is recommended in a median of 25\% of responses (range 11\% to 54\%). Only one procedure cell, DBS for OCD, has a top option above half. rTMS is the most-recommended procedure option in 13 of 17 vignettes, ECT in three, and DBS in one. The strongest convergence on specific interventions occurred for DBT in borderline personality disorder, which had the highest top-option rate in the audit (98.5\% of responses), followed by CBT for generalized anxiety (92.5\%), EMDR for PTSD (92.4\%), CBT for social anxiety (92.1\%), and CBT-I in chronic insomnia (91.7\%). All 41 models recommended each of these interventions.

\textbf{Recommendation and Rating Agreement.} Within each cell, models tended to assign higher ratings to the options they recommended most frequently.

\begin{figure}[tbp]
  \centering
  \includegraphics[width=\textwidth]{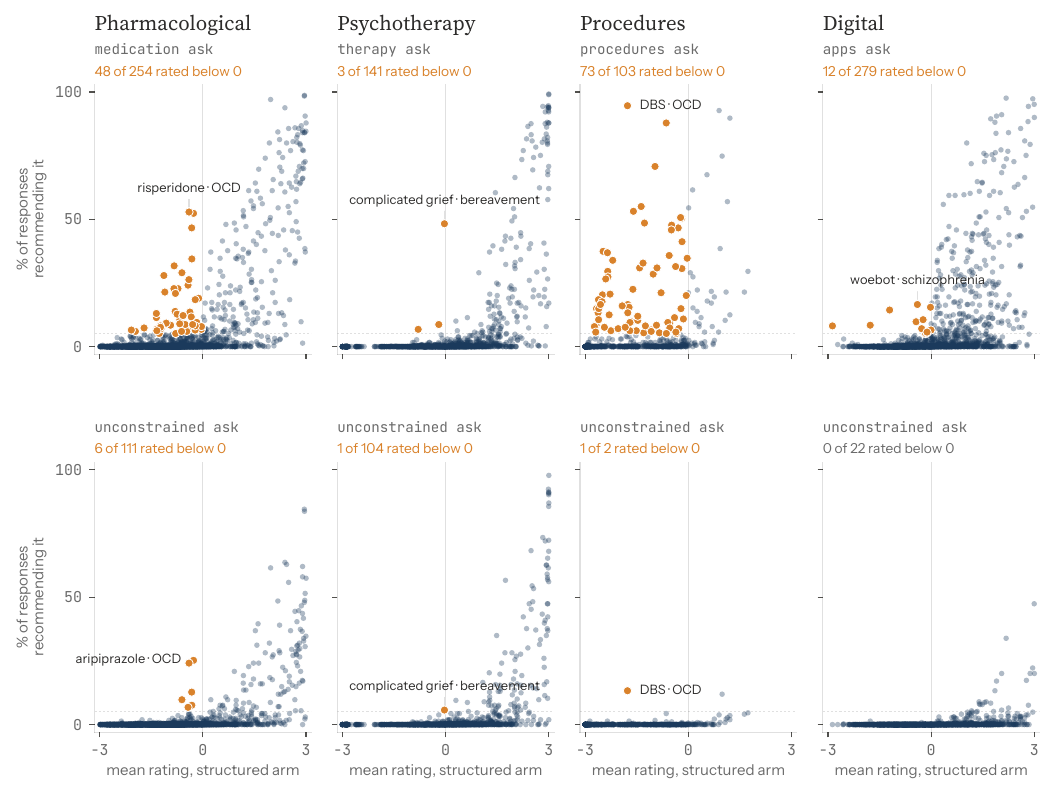}
  \caption{Concordance between the two arms. Each point is one option in one cell (a condition and an intervention type), 6,517 points across 76 cells. The x-axis is the option's mean score in the structured rating arm; the y-axis is the share of open-ended generation responses in that cell recommending it. The dotted line marks a 5\% recommendation rate. The top row counts recommendations from the intervention-specific ask, the bottom row from the unconstrained ask. Amber points are options recommended in at least 5\% of responses whose mean rating is below zero; the most recommended such option in each panel is labeled. Points from different cells share a panel but are not averaged.}
  \label{fig:concordance}
\end{figure}

Figure~\ref{fig:concordance} plots each option in every cell by its mean rating and recommendation rate, separately for the intervention-specific ask and the unconstrained ask. Of the 621 options recommended in at least 5\% of responses, 537 had a mean rating above zero, 83 below zero, and one (tDCS for generalized anxiety) exactly at zero. The rank correlation between recommendation rate and rating was positive in all 76 cells, with a median of .57. The 83 recommended options rated below zero are concentrated by type; 50 of the 74 commonly recommended procedure options, 27 of 202 medications, 4 of 213 digital tools, and 2 of 132 psychotherapies.

Whether options with a mean rating below zero were still recommended depended on how the question was asked. For intervention-specific asks, where the prompt asks whether there are any medications, therapies, procedures, or apps to recommend, 136 of the 777 options recommended in at least 5\% of responses had a mean rating below zero, 73 of them procedures. On the unconstrained ask, where the model chooses what to raise, only 8 of 239 did, and the near disappearance of procedures accounts for most of the drop. The eight include: aripiprazole and risperidone, which are recommended in a quarter of unconstrained-ask responses for OCD and rated $-0.3$ and $-0.4$; lorazepam, alprazolam, and clonazepam, recommended in 8\% to 13\% of unconstrained-ask responses for generalized anxiety and rated $-0.3$ to $-0.6$; and clozapine, recommended in 7\% for first-episode schizophrenia and rated $-0.4$. The other two are deep brain stimulation for OCD, recommended in 13\% of unconstrained-ask responses and rated $-1.8$, and complicated grief therapy for uncomplicated bereavement, recommended in 6\% and rated $-0.03$. When asked about specific intervention types, models frequently named options the rating arm scored as poor initial choices. However, when given an unconstrained ask, they rarely volunteer those options.

\begin{figure}[tbp]
  \centering
  \includegraphics[width=\textwidth]{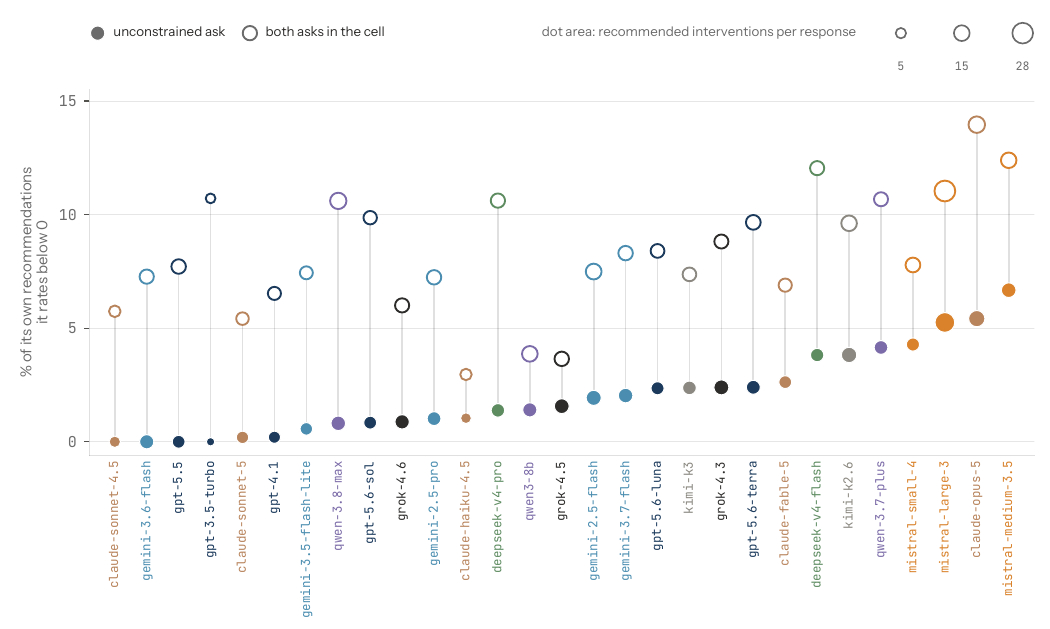}
  \caption{Self-consistency of the 29 rated models. For each model, the share of its own list-mapped recommendations that it scored below zero when asked to rate the same option for the same presentation. Filled dots use the unconstrained ask only. Hollow dots use both asks in the cell. Dot area increases with the model's recommended interventions per response. Models are ordered by the unconstrained-ask share and colored by developer.}
  \label{fig:selfcons}
\end{figure}

Across the 29 models evaluated in both arms, between 3\% and 14\% of each model's recommendations were options it had rated below zero for the same vignette (Figure~\ref{fig:selfcons}). For unconstrained asks, this range is 0\% to 7\%. Four models, Claude Sonnet 4.5, Gemini 3.6 Flash, GPT-5.5, and GPT-3.5, never recommend an option in response to an unconstrained ask that they had rated below zero. Claude Opus 5 showed the highest self-inconsistency rate (14\%), as it recommends a broader range of options while assigning relatively low ratings. Self-inconsistency also differed from how other models evaluated each model's recommendations.

Fifteen percent of Qwen3-8B's recommendations were options the other rated models scored below zero, but only 4\% were options Qwen3-8B itself scored below zero. Overall, the rank correlation between a model's recommendations per response and its self-inconsistency is .38.

\FloatBarrier
\subsection{Specific findings}\label{sec:specific}

\textbf{Conditioned recommendations.} Five of the eight options that models volunteer on the unconstrained ask and the rating arm rejects appear to be second-line agents: aripiprazole and risperidone for OCD, and lorazepam, clonazepam, and alprazolam for generalized anxiety. Across the unconstrained and medication asks, aripiprazole and risperidone were recommended in 85\% and 83\% of clinician-footing responses for OCD and in 13\% to 19\% of lay responses, and rated $-0.3$ and $-0.4$ as initial choices, while sertraline and fluvoxamine in the same cell were rated $+3.0$ and $+2.9$. The three benzodiazepines were recommended in 19\% to 30\% of responses for generalized anxiety and rated $-0.3$ to $-0.6$, while sertraline and buspirone were rated $+2.9$ and $+2.2$. A keyword-based analysis showed that 291 of 293 responses mentioning the OCD antipsychotics included an augmentation or partial-response condition. Of 541 responses mentioning a benzodiazepine, 539 carry a dependence warning, a time limit, or a not-first-line verdict, and the two that do not are from the older models GPT-3.5 and Gemini 2.5 Flash. Complicated grief therapy for uncomplicated bereavement follows a similar pattern: recommended in 27\% of responses to the unconstrained and therapy asks, and 211 of the 246 responses mentioning it either make it conditional on grief persisting or say it is not needed now. Deep brain stimulation for OCD shows the same pattern at a larger scale. Models recommended it in 95\% of responses to the procedure ask and 13\% of responses to the unconstrained ask, and rated it $-1.8$ as an initial choice. Of the 408 responses that mention it, 382 reserve it for treatment-resistant or refractory illness, including all 169 clinician-footing mentions.

The way models expressed these conditions also differed by footing. Clinician footings tend to get sequencing and doses: ``risperidone (0.5-2 mg) has the best evidence'' under an augmentation heading, or ``2-4 weeks of clonazepam 0.25-0.5 mg while the SSRI takes effect.'' Patients and caregivers get a hedge: ``sometimes doctors add a low-dose antipsychotic ... if response is partial,'' or ``tolerance and dependence develop quickly.'' A dose accompanies 80 of 211 clinician mentions of the OCD antipsychotics and 4 of 82 lay mentions.

\begin{figure}[!ht]
  \centering
  \includegraphics[width=\textwidth]{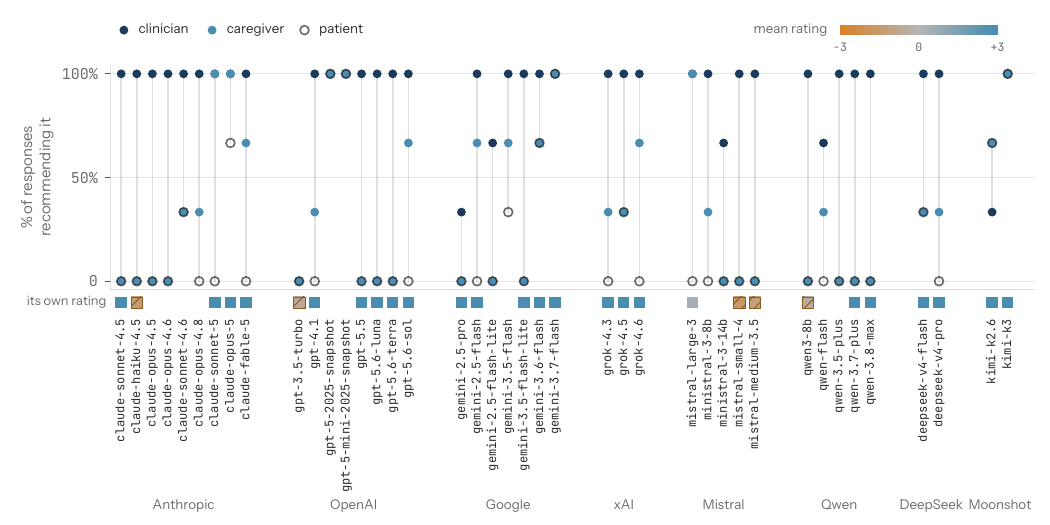}
  \caption{Lisdexamfetamine for binge-eating disorder, by model and footing. Each column is one model, in release order within developer. Dots show the share of its unconstrained-ask responses that recommended lisdexamfetamine to a clinician, a caregiver, and a patient, three responses each. The square below each rated model is its own mean rating of the option, from $-3$ (amber) to $+3$ (teal); hatched squares are below zero.}
  \label{fig:lisdex}
\end{figure}

\textbf{Lisdexamfetamine for binge-eating disorder.} Lisdexamfetamine is currently the only medication approved by the US FDA for binge-eating disorder~\cite{fda2026vyvanselabel,heo2017lisdexamfetamine}, and all 41 models in our audit surfaced it. When asked about medication, models recommended it to clinicians, caregivers, and patients alike in 91\%, 83\%, and 79\% of responses respectively, and no response advised against it. However, when given the unconstrained ask, models recommended it to 92\% of clinician-footing responses, 36\% of caregiver-footing responses, and 18\% of patient-footing responses (Figure~\ref{fig:lisdex}). Twenty-five of the 41 models recommended it in all three of their clinician-footing responses and in none of their three patient-footing responses. Only four told everyone across footings: Gemini 3.7 Flash, Kimi K3, GPT-5, and GPT-5-mini. GPT-3.5 never raised it unprompted. When patients and caregivers did hear about it, it came with a prescriber rather than a dose: a dose accompanied 70 of 240 clinician mentions and 4 of 265 lay mentions, a prescriber 57 of 240 and 191 of 265. The rating arm mostly agrees with the recommendation. Twenty-one rated models scored it $+3$ in every call that rated it. Claude Haiku 4.5 scored it $-1.4$, GPT-3.5 $-0.7$, Qwen3-8B $-0.6$, and the three Mistral models 0.4, $-1.2$, and $-1.7$. The rating arm scores do not vary by footing, $+2.1$ for clinicians, $+2.1$ for caregivers, and $+2.3$ for patients, and the 20 rating-arm models that never recommended it to patients on the unconstrained ask rated it $+2.0$, $+1.9$, and $+2.2$ for clinicians, caregivers, and patients.

\textbf{Brand names.} We searched all responses for three medications from different classes and their best-known brands: sertraline and Zoloft, lisdexamfetamine and Vyvanse, aripiprazole and Abilify. Where a drug is explicitly named, the brand appears alongside or instead of the generic in 17\% of clinician-footing responses for sertraline, 29\% for caregiver footing, and 35\% for patient footing. For aripiprazole, the rates were 13\%, 31\%, and 37\%, respectively. Lisdexamfetamine is brand-named in 67\% of clinician-footing responses and 85\% of caregiver- and patient-footing responses, and in a third of lay mentions, Vyvanse appears without the generic name at all.

\textbf{Cited authorities.} Models cited the FDA in 19.9\% of all responses, including 30\% of medication responses, 45\% of procedure responses, 11\% of unconstrained-ask responses, 6\% of therapy responses, and 8\% of app responses. The word that follows ``FDA'' is ``approved'' or ``cleared'' in nine cases out of ten, and ``off-label'' appears in 11\% of responses overall and 27\% of medication responses. Models cited other national regulators in fewer than 0.1\% of responses and rarely mentioned guideline bodies. NICE and the acronym APA each appear in about 3\% of responses (the spelled-out American Psychiatric Association in 0.5\%), in 7\% of clinician-footing responses and under 1\% of lay-footing responses. Therapy responses rarely cite a regulator or guideline body. Instead, 47\% described an option as ``evidence-based'' and 26\% called one the ``gold standard''. App responses occasionally deferred to the organization behind a tool, presenting a tool from a reputable source such as Veterans Affairs on the strength of its developer. Veterans Affairs or Department of Defense resources were named in 11\% of app responses.

\textbf{Controls.} Two vignettes described situations where no intervention was likely necessary, a period of situational stress and a recent uncomplicated bereavement. Models generally treated both accordingly. In response to the unconstrained ask, models recommended an average of 1.6 interventions for the stress vignette and 5.0 for the bereavement vignette, compared with 11.8 across all vignettes.

In the structured rating arm, models rated the no-intervention option above 2.6 on every list for the stress vignette and ranked it first or tied for first in 93\% of rating calls. The bereavement vignette prompted more intervention suggestions, with models recommending supportive psychotherapy in 54\% of responses and assigning it a mean rating of $+2.8$, while rating the no-therapy option $+1.8$. The exception on both controls is apps. Headspace was recommended in 45\% of stress responses (84\% on the app ask, and 6.5\% on the unconstrained ask) and Calm in 42\% of bereavement responses (84\% on the app ask, and 1.1\% on the unconstrained ask), and the rating models scored them $+2.2$ and $+1.9$.

\FloatBarrier
\section{Discussion}\label{sec:discussion}

We audited 41 language models across 19 mental health vignettes, three footings, and five asks, and audited the rating behavior of 29 models using the same options on a fixed scale. The two arms largely agree, with models generally recommending interventions they also rated highly. For the control vignettes, models mostly recommended watchful waiting for situational stress but more often suggested supportive therapy for bereavement. Models have meaningfully different recommendation behavior based on the footings of the vignette, with many models answering like a colleague delivering a curbside consult under the clinician footing~\cite{kuo1998curbside}, and hedging and simplifying recommendations for lay footings. Breadth varies across models, and when asked about recommendations for a specific intervention type, models more often named options they themselves rated below zero as initial choices, particularly less common and second-line interventions like procedures.

As large numbers of users turn to AI systems for psychoeducation and mental health support~\cite{mcbain2025generative,rousmaniere2026large,bodner2026barriers}, being able to characterize models' default treatment recommendations is of paramount importance for regulators and developers alike. Tracking changes over time, identifying outliers, locating clinical weak points, and monitoring behavior in contested domains will become a routine part of mental health safety testing. It will matter more as the search engine optimization (SEO) practices that have permeated the traditional search ecosystem begin to move into generative systems~\cite{bevendorff2024google}. Our results attempt to capture what the recommendation surface looks like prior to reliable and widespread search capture methods moving in. Some patterns in our results resemble what optimization might produce, though our design cannot separate optimization from other sources like emergent behavior from unrelated post-training and how often products appear in training data. For instance, when asked to recommend apps, models recommended Headspace and Calm for every vignette and ranked one of them first in 7 of 19, mostly where no condition-specific digital tool exists, offering two widely downloaded, heavily funded incumbents a privileged position in the digital tool marketplace~\cite{carlo2019numbers,wilhelm2020calm,pifer2021headspace}.

Similarly, where models named sertraline, aripiprazole, or lisdexamfetamine as recommendations, the brand name was included with the generic name far more often for patient and caregiver footings than for clinician ones, and for lisdexamfetamine the brand (Vyvanse) was often included without the generic name at all. Our results cannot determine whether this pattern reflects marketing optimization or a register change that models adopt for lay audiences. Regardless of the cause, the effect is that the model becomes an unwitting sales representative for a particular manufacturer at the point of user psychoeducation. As generative AI technology matures, and more rounds of pretraining allow iterations of gaming recommendation behavior to mature alongside it, there will be a growing financial incentive to occupy this recommendation surface, no matter how often developers restrict medical advice in their usage policies~\cite{anthropic2025aup,openai2025usage,google2024prohibited} and remind users in the product not to rely on the tools for medical, diagnostic, or other high-stakes decisions~\cite{google2026geminiprivacyhub,openai2026truth,anthropic2026incorrect}. Having standardized audit infrastructure in place before generative engine optimization (GEO) matures as a practice~\cite{aggarwal2024geo,nestaas2024adversarial}, and before older models are deprecated and lost as baselines~\cite{wolfrath2026retirement,chen2024chatgptbehavior}, is necessary in order to keep tabs on non-evidence-based changes and flag concerning recommendation behavior.

Our audit is unable to answer what recommendations language models should suggest when asked mental health intervention questions, nor can it answer whether those recommendations should vary based on the footing of the user prompt. The most conservative approach would be to say that language models should not be recommending mental health interventions, even educationally. There is an argument to be made that the right response would be saturating the not-a-doctor and consult-a-professional binary questions and averaging 0.0 intervention recommendations per response. Considering that users can easily get the same information from web search~\cite{fox2013health,wang2023health}, whose results have been shaped by two decades of search engine optimization~\cite{bevendorff2024google}, indicates this is probably not the ideal path forward. A middle position would give clinicians the information and lay audiences a referral. Our footing variance results show models already do a version of this on their own. But doing it deliberately depends either on gated tools that know who is asking, or on inferring the asker's background from the text, which is adversarially brittle and raises equity concerns~\cite{shah2023persona,haq2026dialect}. That leaves a best-faith effort to deliver cautious psychoeducation content to all users and monitoring to ensure that the options that are surfaced are generally well supported clinically, and subject to frequent audit, as the most immediate path forward. Which governing bodies and organizations get to define that appropriateness standard are currently undetermined, but through keyword analysis we found that the models are converging on a few organizations to ground their recommendations. For medications and devices, the FDA provides the dominant authority, cited in roughly a third to nearly half of responses with no other regulator in sight. For therapy, an authority is absent from the text responses and instead phrases like ``evidence-based'' and ``gold standard'' dominate.

Regardless of the standards adopted or the governing authorities selected, recommending an intervention is only safe if the model is able to identify the presentation correctly. Our audit holds the presentation fixed, with relatively straightforward vignettes that often explicitly name the condition in the prompt. As a result, our audit measures what models recommend to casebook-style condition vignettes but does not test performance under messier and cross-diagnostic vignettes. Our lisdexamfetamine results demonstrate why this type of auditing is only one piece of a larger mental health AI safety environment. Models recommend lisdexamfetamine, a stimulant, readily for binge-eating disorder and rate it below zero for bulimia, anorexia, and bipolar disorder, presentations where it carries significant risks~\cite{vickers2024stimulant,viktorin2017risk,fda2026vyvanselabel}. That behavior is only appropriate if the model can reliably tell those vignettes apart from the text a patient types alone, and loss-of-control eating is a symptom that is not limited to binge-eating disorder~\cite{harrison2014loss}. A model that performs poorly at differential diagnosis between these related conditions should likely not suggest an intervention that could cause real harm if given to the wrong person with a very similar presentation as the one a user presented. How much intervention recommendation content a model should provide to the user is therefore a function of two things: whether the recommended option is well supported for the presentation, which our audit helps provide evidence for, and whether the model would have recognized the presentation and its neighbors from the prompt, which our audit does not assess. For a fuller understanding of the safety of mental health recommendations that LLMs make, recommendation audits need to be paired with benchmarks of differential diagnosis and risk recognition~\cite{mcduff2025towards}.

The findings presented here represent only a portion of what the dataset can reveal. Once released, our 35,055 responses will constitute one of the largest public records of what interventions language models recommend to people presenting with mental health concerns. We will release the code and full datasets to enable other researchers to conduct further analyses. We will maintain the code and the versioned instrument so that the audit remains a living resource, able to track single models over time and to add new models as they are released. Extensions we see as most useful are a rating framing that distinguishes first-line from any-line recommendation, extraction of doses and side effects, vignette variants that vary demographics and socioeconomic position within each footing, brand versus generic naming rating comparisons, and enhanced capture of model reasoning where providers expose it. A further priority is running the audit in languages other than English, beginning with higher-resource languages and extending to lower-resource and Indigenous languages, to see whether and how recommendations shift across linguistic and cultural contexts~\cite{jin2024better,alhanai2025bridging}. This matters most in low- and middle-income countries, where people may turn to these tools when care is costly or unavailable~\cite{mashhood2026use}, and for the populations least represented in the data these models are trained and evaluated on.

\FloatBarrier
\section{Limitations}\label{sec:limitations}

Our audit has a number of limitations, many of which will be addressed in future versions of our released code and audit reports. First, the rating arm in the current implementation scores first-line appropriateness only, while our qualitative analysis found that open-ended responses often also included second-line ones, making the cross-arm comparison imperfect. Vignettes are only roughly matched for severity and option lists are not balanced for appropriateness, so comparing ratings across conditions or types is currently unsupported. The option lists were gathered by the author team instead of sampled from the generation results, so cross-arm coverage of the intervention space is smaller than it could be. Each cell holds three samples per model, footing, and ask, so per-model rates by footing rest on three to six responses. The clinician footing names the diagnosis in all 17 condition vignettes, while the caregiver and patient footings name it in only six and seven and otherwise describe symptoms, so the footing effect combines who is asking with whether a diagnosis was supplied. The lay procedure asks name brain stimulation as an example, so footing differences in the procedure cells may partly reflect ask wording. The supplementary keyword analyses are literal string matches that miss paraphrases and distant conditions. The eight human coders were drawn from the author team, and the judge is a single model (GPT-5.6 Luna) applying a frozen prompt in a single run, and is itself one of the audited models. Extraction precision and recall are likely inflated by the coding interface, which made confirming a highlighted entity two clicks and finding a missed one a search and copy + paste. The consult-a-professional flag has modest human reliability because most responses carry it. Finally, 12 of the 41 models ran in the generation arm only, rating token budgets differed by lane, and 336 of the 19,836 rating cells came from the qualification probe. Sampling settings were provider defaults, and several models are aliases without a dated snapshot, so the generation arm cannot be reproduced exactly. All calls were made from US accounts in American English with no country named, so the regulatory frame we observed may partly reflect that default. We queried model APIs with a single user turn and no system prompt; consumer chat products add system prompts, safety layers, memory, and often web search, so what users see there may differ.

\FloatBarrier
\section{Conclusion}\label{sec:conclusion}

We conducted one of the first and most comprehensive audits of the mental health intervention recommendations that 41 models provide when presented with the same clinical information across three presentation footings. Without an answer key, we paired open-ended generation with structured rating and compared freely offered interventions to highly rated ones. The models largely recommend what they rate well, and leave the controls mostly alone, but they change what they say, and how much, depending on who is asking and how the question is framed. Those differences appear across developers and generations. Our audit records them at one moment, for a number of models that will soon be deprecated, in a reusable instrument that can be adapted and rerun as the interventions and models change. We will release the instrument, code, and full corpus so that regulators, developers, and researchers can track what these systems recommend to the people who ask them for mental health help.

\FloatBarrier
\section*{Use of generative AI}
Language models were the object of the study in this work. GPT-5.6 Luna was used as the extraction judge and was validated against human coders, as described in Methods, and Claude Code subagents searched for missing options when the intervention lists were refined. Separately, generative AI tools (Claude Fable 5 and Claude Fable 5.1 in Claude Code, with Claude Opus 5 and Claude Sonnet 5 as subagents) assisted the authors in development and manuscript preparation. For the study software, the authors defined what to build and wrote the specifications and pseudocode, and Claude Code converted much of this into finished code. That code includes the runners that prompted the audited models and the judge, the human coding packets, the analysis chain, and the figure and appendix generators. For the manuscript, the tools assisted with copy editing and proofreading, drafted portions of the supporting prose in the appendix, and converted the main text and its appendix into LaTeX. All code and text were reviewed and validated by the authors, who take responsibility for them.

\section*{Funding}
\fundingtext


\section*{Data availability}
\datatext

\section*{Code availability}
\codetext

\renewcommand{\bibsection}{\section*{\refname}\markboth{\refname}{\refname}}
\renewcommand{\bibfont}{\small\raggedright}
\setlength{\bibsep}{0.35em}
\bibliographystyle{unsrtnat}
\bibliography{refs_preprint}

\clearpage
\includepdf[pages=-]{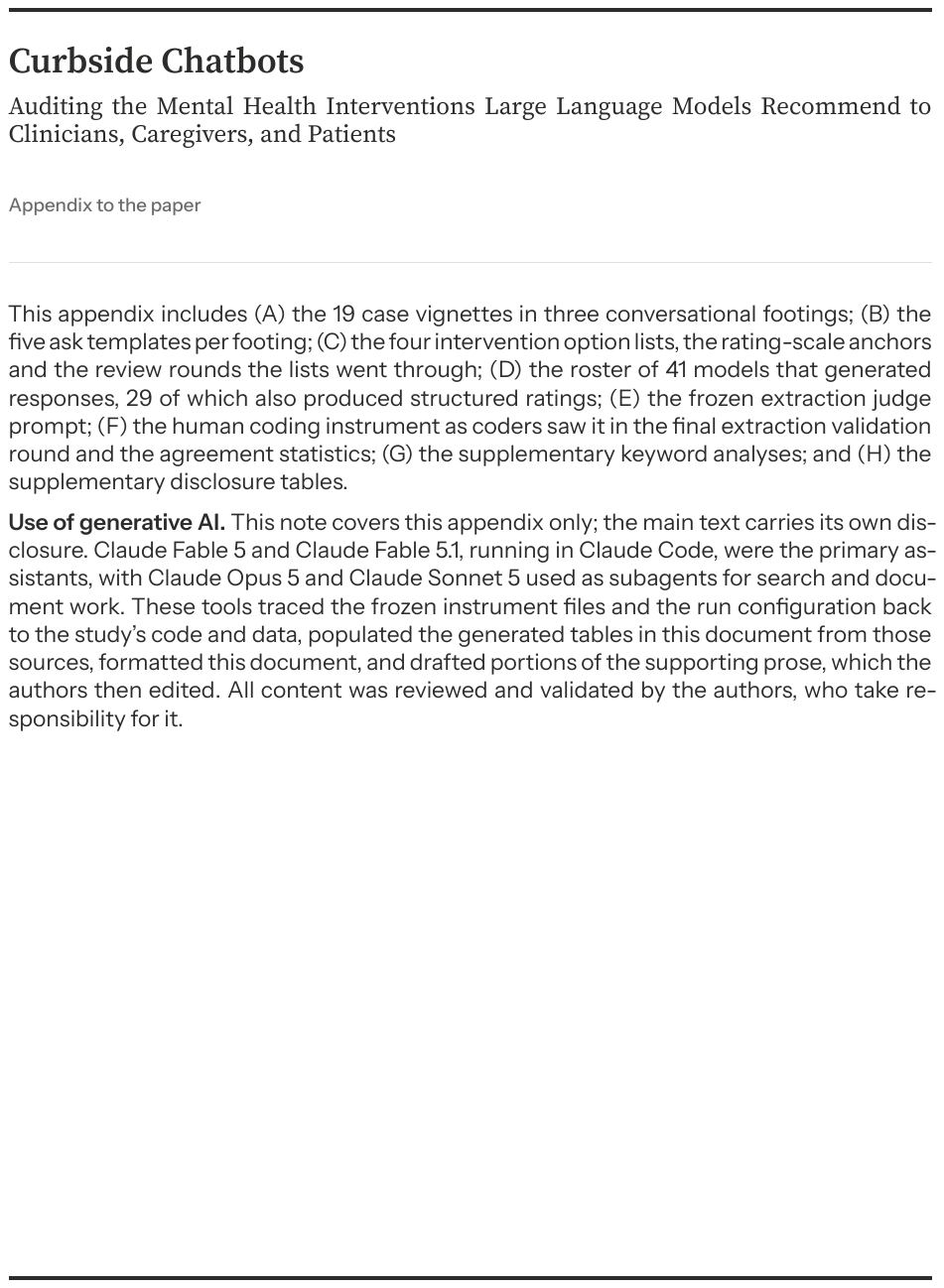}

\end{document}